\documentclass[12pt,oneside,english]{amsart}
\usepackage[T1]{fontenc}
\usepackage[latin9]{inputenc}
\usepackage{amsthm}
\usepackage{amstext}
\usepackage{amssymb}
\usepackage{graphicx}

\makeatletter

\providecommand{\tabularnewline}{\\}

\numberwithin{equation}{section}
\numberwithin{figure}{section}
  \theoremstyle{remark}
  \newtheorem*{acknowledgement*}{\protect\acknowledgementname}
\theoremstyle{plain}
\newtheorem{thm}{\protect\theoremname}
  \theoremstyle{definition}
  \newtheorem{example}[thm]{\protect\examplename}
  \theoremstyle{plain}
  \newtheorem{lem}[thm]{\protect\lemmaname}
  \theoremstyle{plain}
  \newtheorem*{cor*}{\protect\corollaryname}
  \theoremstyle{remark}
  \newtheorem{note}[thm]{\protect\notename}
  \theoremstyle{remark}
  \newtheorem{rem}[thm]{\protect\remarkname}
  \theoremstyle{plain}
  \newtheorem*{thm*}{\protect\theoremname}

\makeatother

\usepackage{babel}
  \providecommand{\acknowledgementname}{Acknowledgement}
  \providecommand{\corollaryname}{Corollary}
  \providecommand{\examplename}{Example}
  \providecommand{\lemmaname}{Lemma}
  \providecommand{\notename}{Note}
  \providecommand{\remarkname}{Remark}
  \providecommand{\theoremname}{Theorem}
\providecommand{\theoremname}{Theorem}

\begin{document}

\title{Short Title: Theory on Reporting}

\maketitle

\section*{Understanding Theoretically The Impact of Reporting of Disease Cases
in Epidemiology}

\vspace{0.1cm}

To appear in \emph{Journal of Theoretical Biology (Elsevier)}

\vspace{0.5cm}

\begin{center}

\author{\textbf{ARNI S.R. SRINIVASA RAO}}

Bayesian and Interdisciplinary Research Unit,

Indian Statistical Institute, 

203 B.T. Road, Calcutta, INDIA 700108. 

Email: arni@isical.ac.in. 

Tel: +91-33-25753511.

\end{center}

\vspace{0.5cm}

\textbf{AMS subject classifications:} 92D30, 26.70\vspace{0.2cm}
\begin{abstract}
In conducting preliminary analysis during an epidemic, data on reported
disease cases offer key information in guiding the direction to the
in-depth analysis. Models for growth and transmission dynamics are
heavily dependent on preliminary analysis results. When a particular
disease case is reported more than once or alternatively is never
reported or detected in the population, then in such a situation,
there is a possibility of existence of multiple reporting or under
reporting in the population. In this work, a theoretical approach
for studying reporting error in epidemiology is explored. The upper
bound for the error that arises due to multiple reporting is higher
than that which arises due to under reporting. Numerical examples
are provided to support the arguments. This article mainly treats
reporting error as deterministic and one can explore a stochastic
model for the same.
\end{abstract}

\keywords{Key words: Multiple reporting, diagnosis, adjustment.\vspace{0.5cm}}
\begin{acknowledgement*}
My gratitude to Professor Lord Robert M. May of Oxford for his excellent
encouragement and appreciation to continue this work when I had shown
initial draft. Part of the work was done during 2005-2007 while I
was at Center for Mathematical Biology, Mathematical Institute, University
of Oxford. I am grateful to the referees for their several helpful
suggestions, comments to revise and making very generous comments
about the work.
\end{acknowledgement*}
\pagebreak

\section{Introduction}

Reporting is one of the crucial elements of epidemiological research.
Its importance ranges from helping the base line assessment of the
epidemic to understanding the rate of reproduction of infected individuals.
For example, a simple equation of the form $I(t)=I(0)\exp(a.t)$ can
be used to estimate $a,$ the exponential growth rate between the
reported infection numbers $I(0)$ and $I(t)$ at times $0$ and $t$
($t>0)$ respectively. When $I(0)$ and $I(t)$ suffer with reporting
errors or when they lack accuracy, then the computed growth rate $a$
is misleading. There are evidences that under reporting of the cases
lead to under estimation of incidence \cite{Alter1987,Mubayi2010},
delay in monitoring and surveillance \cite{Jelast2010,Jelast2009}.
There are studies which support better idea on the magnitude of the
epidemic had there been no under reporting \cite{Lipsitch2009,Jorgen2008}.
Since under reporting could mislead the impact of the epidemic, there
were attempts to understand the extent of under reporting using various
surveys and modeling \cite{Singh2006,Brum2005,Hest2002,Bernillion2000}.
There are several deterministic and stochastic models available for
computing the growth rates of epidemics, see \cite{AM1991,BPW2008}.
There are certain methods which fail to predict epidemic growth accurately
or fail to ascertain the past trends of the infections when reporting
is incomplete. The method of back-calculation \cite{BaG1988} for
estimating HIV infection fails to construct HIV trends accurately
when AIDS reporting is incomplete.\textbf{ }Such methods are based
on the fact that, after assessing the number of individuals with infection,
the duration between infection times and disease times is used to
project number of individuals with disease at some future time point.
Here, instead of handling the cases discretely, convolution of infection
density and density of duration between infection and disease times
for relevant continuous random variables are considered. Future numbers
of individuals with disease already projected using back-calculation
methods can be compared with number of reported disease cases at the
same point to obtain reporting error of disease cases. By application
of such methods, it is implicitly assumed that the populations are
closed to migration during the study period or during the two time
points where a reporting error of disease is estimated. Other popular
methods for reporting error include, conducting surveys at two or
more time points on a population which involve either testing of randomly
selected blood samples for infection under study or assessing infected
people through verbal autopsies and then comparing the estimated infection
prevalence in the same population with already existed reported infection
numbers at the same time.\textbf{ }In general, for simple or advanced
models, if data suffers from under-reporting then usually the data
is adjusted before applying a given method. Reported incidence and
prevalence are requirements for validating models and forecasting.
Also, the parameters derived from these reported incidence trends
are shown to be consistent in model building and analysis \cite{AM1991}. 

Over media coverage of Swine Flu in some parts of the world led to
over magnifying of the disease burden as these preliminary results
were used in modeling epidemics in many countries during 2009 outbreak
of novel H1N1 influenza. It could have happened that in the 2009 swine
flu outbreak, some studies disregarded the large number of cases that
did not lead to any serious complications. Protocols and preparedness
for future pandemic based on the experience of 2009 outbreak in Europe
is well understood \cite{Nicoll2010}. In a study on BSE (Bovine Spongiform
Encephalopathy) in France, it was found that some cases were not detected
by the surveillance system, which caused under reporting of the epidemic
\cite{Supervie2004}. In this study, they reconstructed the past trends
by back-calculation and adjusted the under reporting. Another study
on BSE in Britain examined the under reporting of cases and differential
mortality using back-calculation by improving the standard back-calculation
technique \cite{Donnely2002}. Measles data analysis in Italy indicated
that under reporting could be distorting observed epidemic patterns
\cite{Williams2003}. A study \cite{Boerma2006} on HIV addresses
that over reporting of individuals on antiretroviral therapy and related
caution to be taken while estimating the number. Over reporting percentage
was found to be important to ascertain actual epidemic levels in sexually
transmitted infections in Amsterdam \cite{Fennema1995}.

There are several ways of quantifying the reporting error depending
upon the epidemic. These could be observing incidence curve obtained
by models with reported incidence of a given epidemic, through sample
surveys, back-calculation methods etc, for example, see\textbf{ \cite{Brum2005,Hest2002,BaG1988}}.
In this paper, a supplementary way is proposed for understanding efficiency
of reporting using limit analysis. In this context, the terms 'limit
analysis' meant that the rate of increasing or decreasing of reporting
efficiencies are studied over a very long period of time, and also
situations such as reported number of disease cases approaching to
actual disease cases are studied while obtaining bounds of error.
Numerical examples are also provided. Our method treats reported and
actual disease cases as numbers on the real line and functions of
error of reporting are proposed to quantify the bounds of error of
reporting. We introduce theoretical arguments in different settings
and illustrate them by numerical examples. The upper bound for this
calculated error that arises due to multiple reporting (excess reporting)
is shown here to be lower than that of error due to under reporting.
We also analytically show that even if error of reporting is not observed,
there is a possibility of multiple reporting in the data. Realistic
data fitting is not in the scope of present work. The results indicate
that there exists a serious consequence to multiple reporting (a situation
arises when each case is reported more than once).

\section{Preliminaries}

Reporting of disease cases plays an important role in understanding
epidemics. We provide two examples and two observations.
\begin{example}
\label{exa:1}Consider a homogenous population of 800 individuals,
where each individual has equal chance of acquiring an infection of
type A virus. Suppose $7$ individuals were reported of acquiring
infection of type $A$ by the health system in a year of $26$ actual
number of cases infected in the same year. Now, the prevalence of
type $A$ virus in this year is $7/800=0.00875$, but actual prevalence
after adjusting for under reporting is $26/800=0.0325$. Percentage
of reported out of actual cases in this situation is 26.92. 
\end{example}
$ $
\begin{example}
\label{exa:2}Let us now compute incidence rate of type $B$ virus
in a cohort study. Suppose a cohort of $750$ individuals are followed
for one year and during which $17$ new cases were reported in the
year to have acquired type $B$ virus out of $48$ actually acquired
the virus in the same year. The incidence rate by assuming uniform
distribution of infections over the year is $17/741.5=0.0229$ person-years,
where as actual incidence rate after adjusting for under reporting
is $48/726=0.0661$ person-years. Note that each of the 17 reported
cases were remained uninfected on an average of six months, hence
750 individuals were actually followed for 750-8.5=741.5 years without
being infected in that year. By a similar explanation for 48 actual
cases, we obtain 726 person-years. In an ideal situation, well designed
cohort studies consists at least information on number of individuals
recruited for the study, duration of follow-up for each individual
and number of newly infected cases of virus during the study period.
Among other reasons, under reporting could arise also due to both
infection and recovery from the virus between two follow-up periods
and not detecting the virus at the time of the next follow-up, not
reporting at the time of verbal autopsy conducted at next follow-up
where clinical diagnosis for the presence of the virus were conducted
etc. 
\end{example}
The under reporting or over reporting of cases leads to errors in
assessing the epidemic spread through modeling. Total disease cases
(i.e. the number of actual cases) in the population could be taken
as the reported number plus or minus the error of reporting. In the
present work, it is attempted to study when efficiency in reporting
error is considered as a difference between $\Lambda_{h}$(number
of total cases at time $h$) and $\Omega_{h}$ (number of reported
cases at time $h$). The three situations that arise are, i) $\Lambda_{h}$>
$\Omega_{h}$ (due to under reporting), ii) $\Lambda_{h}$< $\Omega_{h}$
(due to over reporting) and iii) $\Lambda_{h}$= $\Omega_{h}$(due
to accurate reporting or due to no reporting error\textbf{, }when
there are no multiple reported cases among reported cases\textbf{). }

\emph{Observation 1.} We saw from the examples \ref{exa:1} and \ref{exa:2},
that there is no error (or some may term it as no bias) in estimating
incidence or prevalence when the\textbf{ }ratio $\Omega_{h}/\Lambda_{h}$
attains the value $1.$ We define neighbourhood around actual cases
$\Lambda_{h}$ for some $\sigma>0$ be\textbf{ $\mathbf{B}_{\sigma}(\Lambda_{h})=\left\{ b\in\mathbb{R}:\left|b-\Lambda_{h}\right|<\sigma\right\} $
}and define neighbourhood around $1$ for some $\omega>0$ be $\mathbf{A}_{\omega}(1)=\left\{ a\in\mathbb{R}:\left|a-1\right|<\omega\right\} .$\textbf{
}Then for every $\mathbf{A}_{\omega}(1)$, there exists a $\mathbf{B}_{\sigma}(\Lambda_{h})$
with the property that for all $\Omega_{h}\in\mathbf{B}_{\sigma}(\Lambda_{h}),$
it follows that $\Omega_{h}/\Lambda_{h}\in\mathbf{A}_{\omega}(1).$\textbf{
}In the next section, we argue that $(\Omega_{h})$ is bounded. By
adopting results in \cite{Copson70/71} to the present epidemiology
scenario, we can deduce that $(\Omega_{h})$ is convergent when $(\Omega_{h})$
is bounded (if we obtain the inequality $2\Omega_{h+2}\leqslant\Omega_{h+1}+\Omega_{h}).$
Further, under a certain assumption, we see that $(\Omega_{h})$ is
convergent without above inequality. The fact that the above type
of inequality is not necessary for a bounded sequence to convergent
was discussed with an example in \cite{Copson70/71}. 

\emph{Observation 2.} Let $x_{h}$ be a random variable such that
$x_{h}\in(0,1)$. If $\lambda_{m}=\Lambda_{1}+\Lambda_{2}+\Lambda_{3}+...+\Lambda_{m}$
and $\omega_{m}=\Omega_{1}+\Omega_{2}+\Omega_{3}+...+\Omega_{m},$
then, the following were observed \cite{Rao2004}:

\textbf{
\begin{align*}
i)\:\lambda_{m} & >\omega_{m}\\
\\
ii)\:\lambda_{m} & =\sum_{h=1}^{m}\sum_{k=0}^{\infty}\Omega_{h}x_{h}^{k}.
\end{align*}
}

Further, when $\bold{\Omega}$ follows Poisson mass function with
parameter $P$ and rate of decrease of ${x_{h}'s}$ is $c$, it was
observed \cite{Rao2004} that

\textbf{
\begin{eqnarray*}
\left(\frac{\exp\{-P\}P^{\Omega_{h}}}{\Omega_{h}!},\frac{x_{0}\exp\{-(P+c.h)\}P^{\Omega_{h}}}{\Omega_{h}!},\frac{x_{0}^{2}\exp\{-(P+2c.h)\}P^{\Omega_{h}}}{\Omega_{h}!},...\right)
\end{eqnarray*}
}is convergent. 

Multiple reporting phenomena might also contribute in reduction of
efficiency in reported cases. In this work, efficiency is not only
measured as a difference of reported and total cases, but also impact
of multiple reporting phenomena is studied. The results presented
here are original and brings a new outlook to study epidemic behavior.

\section{Epidemic reporting efficiency}

We denote, $\alpha_{h}$ for the difference between reported and actual
cases at time $h$. If $\Lambda_{h}$ is total cases, $\Omega_{h}$
is reported cases and $\alpha_{h}$ is error of reporting taken over
the time $h$ then symbolically, $\Lambda_{h}=\Omega_{h}\pm\alpha_{h}$.
As $\alpha_{h}$ tends to zero, then $\Omega_{h}\rightarrow\Lambda_{h}$
for some $h>N\in\mathbb{N}$ (section 3, \cite{Rao2004}), $\Lambda_{h}$
is more than $\Omega_{h}$ (in case of under reporting), $\Lambda_{h}$
is less than $\Omega_{h}$ (in case of multiple reporting) and $\Lambda_{h}$
is equal to $\Omega_{h}$ (in case of no reporting error). There is
some possibility that these under reported cases suffer from multiple
reporting. For instance, let $n_{1h}$ be the number of individuals
out of $\Omega_{h}$ those are reported exactly once, so that $\Omega_{h}-n_{1h}$
is the number those are reported more than once, then $\Omega_{h}=(\Omega_{h}-n_{1h})+n_{1h}.$
This tells us, reported cases need not be of different individuals
and could be sum of those individuals whose cases were reported more
than once and those individuals whose cases reported only once. If
none of the individuals were reported exactly once (a rare event may
arise in case of complete uncertainty of health diagnostics, facilities),
then all the reported cases are sum of multiple reporting cases. If
we denote $f$ for the efficiency of reporting and define it as the
ratio of $\Omega_{h}$ and $\Lambda_{h}$, then $f$ could vary over
the time period depending upon the reporting system. If multiple reporting
is present then, $f(x_{h})=\Lambda_{h}/\Omega_{h}$ and after adjusting
for excess number due to multiple reporting, the resultant efficiency
function will be, $f_{1}(x_{h})=\Lambda_{h}/n_{1h},$ where $n_{1h}<\Omega_{h}.$
Here $f_{1}>f.$ Similarly, $\alpha_{h}=\Lambda_{h}-\Omega_{h}$ or
$\Omega_{h}-\Lambda_{h}$ and $\alpha'_{h}=\Lambda_{h}-n_{1h}$ or
$n_{1h}-\Lambda_{h}$, where $n_{1h}<\Omega_{h}.$ If we assume $\alpha_{h}$
is constant over time (say, $\alpha$) then the difference between
$\Lambda_{h}$ and $\Omega_{h}$ is constant over time $h$. We begin
with elementary case of epidemic efficiency as a difference between
reported and total cases and then extend the case by varying efficiency.

\subsection{\textmd{$\bold{\left(\Lambda_{h}<\Omega_{h}\right)}$}}

This is a situation which raises due to multiple reporting of cases.
The reasons responsible for this are when individuals go to several
clinics or public medical setups to get diagnosis and each of these
clinical or medical setup report to the national level epidemic surveillance.
Individuals may prefer re-diagnosis either due to not having faith
in one particular system where they were detected for a disease or
it could be due to choice of reconfirmation of the diagnosis. Since,
$\alpha>0$, we have $\Omega_{h}-\alpha>0$ and $\Lambda_{h}>0$ $\forall$
$h\in\mathbb{Z^{+}}$. Let us assume that the epidemic grows exponentially
and becomes severe as the time progresses (which is usual in the beginning
for many epidemics), then $(\Omega_{h})$ can be taken as a monotonic
increasing sequence. Let $W$ be the whole population, then $\Omega_{h}\leq CW\,\,\forall\: h$,
where $C\in\mathbb{R^{+}}$ is due to multiple reporting. At any given
point of time, $(\Omega_{h})$ cannot be more than the finite multiples
of the total population. This is because if the epidemic spreads to
entire population and even if each case is reported multiple ways,
still it will be a finite number, i.e $CW$ is finite. Hence $(\Omega_{h})$
is bounded and convergent. Since $\alpha$ is finite then $(\Lambda_{h})$
is also convergent. We have $\Lambda_{h}^{-1}=(\Omega_{h}-\alpha)^{-1}$$=\Omega_{h}^{-1}\left\{ 1-\left(\alpha/\Omega_{h}\right)\right\} ^{-1}$.
From the properties of numbers, whenever $\alpha/\Omega_{h}<1,$ then
we can bring the inequality $1-\left(\alpha/\Omega_{h}\right)^{2}<\exp\left(\alpha/\Omega_{h}\right)(1-\alpha/\Omega_{h})<1.$
This implies $\exp\left(\alpha/\Omega_{h}\right)<(1-\alpha/\Omega_{h})^{-1}$$\Rightarrow(\Omega_{h}^{-1})\exp\left(\alpha/\Omega_{h}\right)<(\Omega_{h})^{-1}\left\{ 1-\left(\alpha/\Omega_{h}\right)\right\} ^{-1}.$
Thus by simplifying we get $\alpha<\Omega_{h}\ln\left(\Omega_{h}/\Lambda_{h}\right)\forall h.$
Let $\widehat{\Omega}$ be the maximum for $\Omega_{h}$ values and
$\widehat{\Lambda}$ be the maximum for $\Lambda_{h}$ values, then
$\widehat{\Omega}\ln\left(\widehat{\Omega}/\widehat{\Lambda}\right)$
can be treated as an upper bound for $\alpha.$ Let $(\Omega_{h})$
be a monotonically non-increasing (and also epidemic does not grow
exponentially), but always maintains the relation $\Omega_{h}-\alpha>0,$
and follows a periodic maximum value with period of $H$ (say) time
points. For this situation also $\widehat{\Omega}\ln\left(\widehat{\Omega}/\widehat{\Lambda}\right)$
is an upper bound for $\alpha$. There is a possibility to have a
smaller upper bound than this for $\alpha.$ Even if $\Omega_{h}$
values stop to behave like periodic maximum property and increase
after some $j>N\in\mathbb{N}$, then $\alpha<\Omega_{h}\ln\left(\Omega_{h}/\Lambda_{h}\right).$
When $\Lambda_{h}\rightarrow0$ then $\Omega_{h}\rightarrow0$. Eventually,
as $\Lambda_{h}\rightarrow0$ then irrespective of the error of the
reporting is high or low, eventually disease cases will become zero,
hence study of $\alpha$ is not considered important in this situation.
Now, we begin with a trivial statement on total reported cases.
\begin{thm}
Let $\epsilon>0.$ If $\Omega_{h}>\Lambda_{h}$, $\Omega_{h}$ is
monotonically increasing function or\textbf{ }monotonic non-increasing
but $\Omega_{h}-\alpha>0$, then there exists a point in the sequence
$(\Lambda_{n})$ such that $\Lambda_{h}\in\mathbf{B}_{\epsilon}(\Omega_{h}),$
where $\mathbf{B}_{\epsilon}(\Omega_{h})$ is $\epsilon-$ neighbourhood
of $\Omega_{h}.$ \end{thm}
\begin{proof}
Let $\epsilon>0$. We have seen in section 3.1 that $\alpha<\Omega_{h}\ln(\Omega_{h}/\Lambda_{h})$
when $(\Omega_{h})$ is monotonically increasing as well $(\Omega_{h})$
is not monotonic increasing but $\Omega_{h}-\alpha>0.$ Therefore,
$\left|\Lambda_{h}-\Omega_{h}\right|<$$\Omega_{h}\ln(\Omega_{h}/\Lambda_{h}).$
When we choose $\Lambda_{h}>\Omega_{h}/\exp(\epsilon/\Omega_{h})$
for some $h>\bold{N},$ then $\Lambda_{h}\in\mathbf{B}_{\epsilon}(\Omega_{h}).$
\end{proof}

\subsection{$\bold{\left(\Lambda_{h}>\Omega_{h}\right)}$}

\begin{figure}
\includegraphics[scale=0.6]{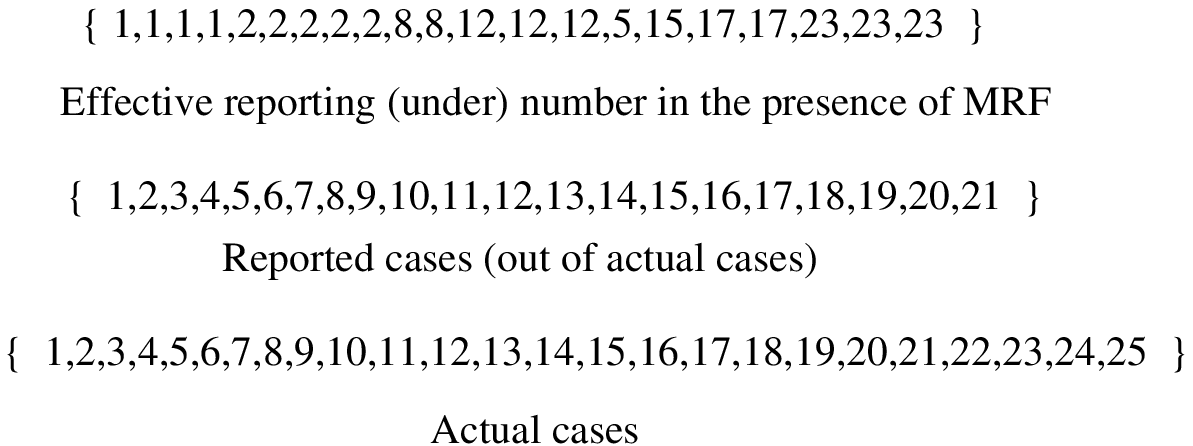}

\caption{\label{figure1}Schematic diagram of 'multiple reporting factor' within
under reporting. In the first row, we observe that $21$ cases are
reported for an epidemic in a certain time period. If we assume there
is no multiple reporting among these $21$ reported cases, we can
consider them as total reported in this period. If we report them
as provided in the second row, then the ratio of reported cases to
the actual disease cases (see third row) is $21/25=0.84.$ However,
observe that, out of $21$ cases reported in the first row, case $1$
is reported $4$ times, case $2$ is reported $5$ times, and so on
case $23$ is reported $3$ times. Removing multiple reported cases
from first row, the number of distinct cases reported are only $8,$
thus the ratio of reported (after adjusting for under reporting) to
actual cases reduces to $8/25=0.32.$ }
\end{figure}

This is a typical under reporting situation which could arise due
to following consequences: incomplete diagnosis, incomplete reporting
of the diagnosed cases and under detection of cases. Here $\Lambda_{h}=\Omega_{h}+\alpha.$
We have $\alpha^{-1}=\left\{ (1/\Lambda_{h})(1-\Omega_{h}/\Lambda_{h})\right\} $
where $\Omega_{h}/\Lambda_{h}<1$. Therefore $1-(\Omega_{h}/\Lambda_{h})^{2}<\exp\left(\Omega_{h}/\Lambda_{h}\right)(1-\Omega_{h}/\Lambda_{h})<1.$
This implies $(1/\Lambda_{h})\exp\left(\Omega_{h}/\Lambda_{h}\right)<$$\Lambda_{h}^{-1}$
$\left\{ 1-\left(\Omega_{h}/\Lambda_{h}\right)\right\} ^{-1}$$=\alpha^{-1}.$
Therefore $\alpha<\Lambda_{h}\exp\left(-\Omega_{h}/\Lambda_{h}\right)$
and $\widehat{\Lambda}\exp\left(-\widehat{\Omega}/\widehat{\Lambda}\right)$
is an upper bound. Even though reported cases are less than that of
actual, there is a possibility of multiple reporting among under reported
cases. Admitting this fact further complicates the error associated
with epidemic analysis. In the presence of such multiple reporting,
under reporting observed is indeed more than that of we normally admit
without taking 'multiple reporting factor' (\emph{MRF}). In other
words, by neglecting \emph{MRF} (when it is present in the data),
the degree of reporting would be better, but it is indeed a false
degree of reporting (Fig. \ref{figure1}). Therefore, MRF within under
reporting implies reporting is further lower than the total cases. 
\begin{thm}
If $\Lambda_{h}>\Omega_{h}$ then there exists a point in the sequence
$(\Lambda_{h})$ such that $\Lambda_{h}\in\mathbf{B}_{\epsilon}(\Omega_{h}),$
for every $\epsilon>0.$ \end{thm}
\begin{proof}
Under the hypothesis, we have $\alpha<\Lambda_{h}\exp\left(\Omega_{h}/\Lambda_{h}\right).$
Therefore, $\left|\Lambda_{h}-\Omega_{h}\right|<\Lambda_{h}\exp\left(-\Omega_{h}/\Lambda_{h}\right)$.
Now, when we choose $\Omega_{h}>\Lambda_{h}(\ln\Omega_{h}/log\epsilon)$
for some $h>\bold{N}$ then $\Lambda_{h}\in\mathbf{B}_{\epsilon}(\Omega_{h}).$
Note that, $\Omega_{h}>\Lambda_{h}(\ln\Omega_{h}/log\epsilon)\Rightarrow$$\epsilon>\Lambda_{h}\exp\left(-\Omega_{h}/\Lambda_{h}\right)$
\end{proof}
$ $

$ $

\subsubsection{Multiple reporting within $\Omega_{h}$ }

\begin{figure}
\includegraphics[scale=0.6]{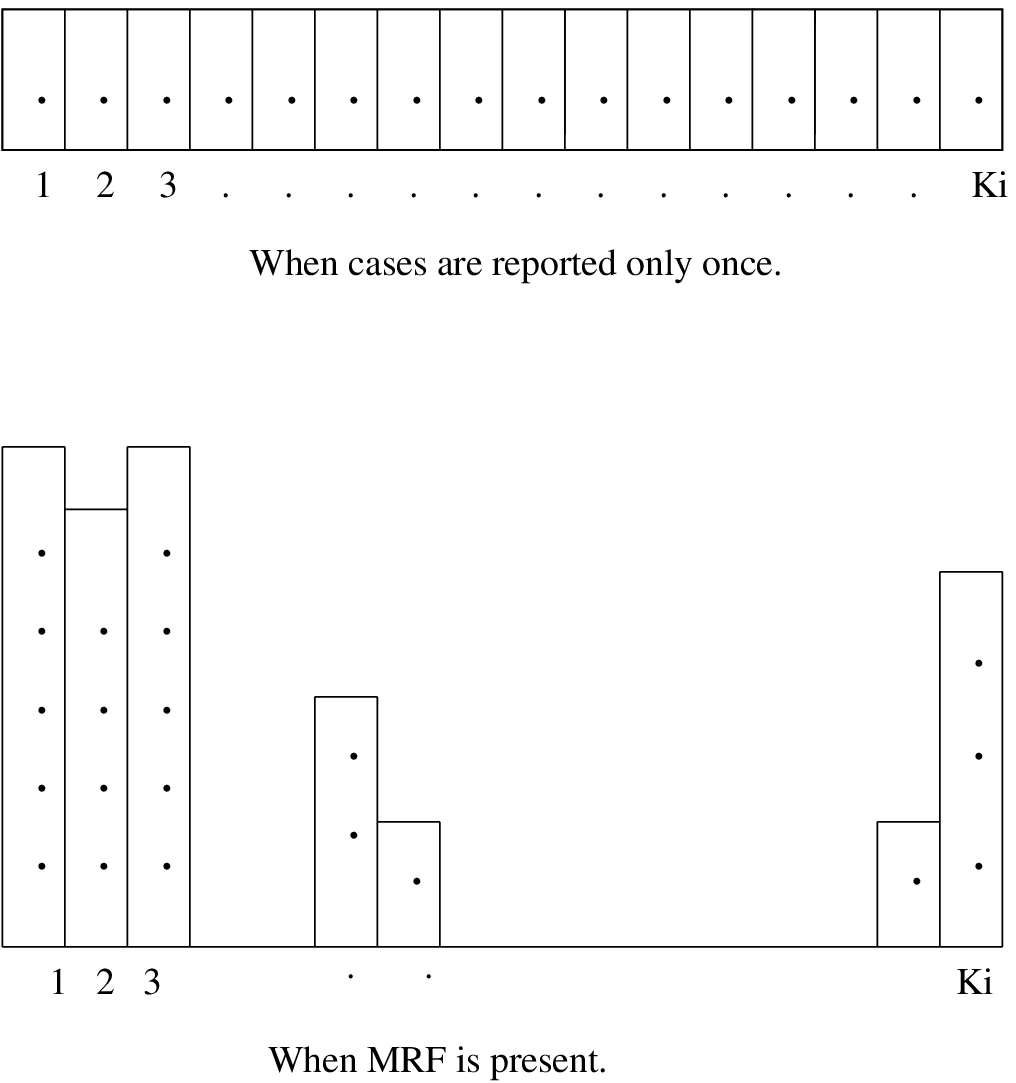}

\caption{\label{figure2}This figures indicates if $\eta_{h}(\in K_{h})$ classes
are empty (i.e. no reported case in these classes), then this is compensated
by more than one reported cases in one or more of the remaining $(K_{h}-\eta_{h})$
classes. }
\end{figure}

Let $K_{h}$ be a positive integer which is defined as number of classes
at time $h$ which can accommodate $\Omega_{h}$. Suppose $\Omega_{h}$
is completely made up of $K_{h}$(say) classes and each class consists
finite number of (multiple) reporting of one individual. If every
class consists of one member then $\Omega_{h}=K_{h}$, a situation
when multiple reporting among reported cases is avoided. On the other
side if $\eta_{h}\,(\in K_{h})$ classes are empty (i.e. no reported
case in these classes), then this is compensated by more than one
reported cases in one or more of the remaining $(K_{h}-\eta_{h})$
classes ($N_{h},$ say) (see also Fig. \ref{figure2}). As $\eta_{h}\rightarrow0$,
the reported cases (under reported number) tends to represent true
(actual) cases and are not affected by multiple reporting of individual
cases. Expected error in the presence of under reporting $\alpha=\Lambda_{h}-(K_{h}-\eta_{h}).$
Even though $\eta_{h}\rightarrow0$, we have to note that actual cases
suffer under reporting. We can observe that $\alpha<\Lambda_{h}\exp\left(-N_{h}/\Lambda_{h}\right)$
and as $\eta_{h}\rightarrow0$ then $\alpha<\Lambda_{h}\exp\left(K_{h}/\Lambda_{h}\right).$
Overall, as $\alpha,\eta_{h}\rightarrow0,$ the reporting error is
minimized and total reported cases is equal to the total (actual)
cases (assuming diagnosis is complete). If $\Omega_{h}\rightarrow\infty$,
then as $\eta_{h}\rightarrow K_{h}$ (or $\eta_{h}$ is high), the
error of reporting is very high. If $\Omega_{h}\approx\textrm{const.},$
then as $\eta_{h}\rightarrow K_{h},$ error of reporting will be still
more than that of expected. When $\Lambda_{h}\rightarrow0$ then as
$\eta_{h}\rightarrow K_{h},$ the error of reporting will decline
too. But this violates the assumption that reporting error is constant.
This condition is out of the scope of this section and we discuss
these issues in the next section. Lower the $\eta_{h}$ implies lower
level of multiple reporting in the population. 
\begin{lem}
\label{lemma1}$\eta_{h}\rightarrow0\Rightarrow\left(\Lambda_{h}-K_{h}\right)\rightarrow\alpha.$\end{lem}
\begin{proof}
We know that $\eta_{h}\rightarrow0\Rightarrow\Omega_{h}\rightarrow K_{h}.$
This means by algebraic limit principle for a given constant $\alpha$,
$\Omega_{h}+\alpha\rightarrow K_{h}+\alpha.$ Therefore $\Lambda_{h}\rightarrow K_{h}+\alpha.$
This implies, for all $\epsilon>0$, there exists an integer $N_{h}$
such that $h>N_{h}\Rightarrow\left|\Lambda_{h}-K_{h}-\alpha\right|<\epsilon$
for some $h$. Therefore $\left(\Lambda_{h}-K_{h}\right)\rightarrow\alpha$.\end{proof}
\begin{cor*}
$\alpha\rightarrow0\Rightarrow\Lambda_{h}\rightarrow K_{h}.$
\end{cor*}
Suppose $\eta_{h}>0$, this means there are some empty classes out
of $K_{h}$ classes, so that $K_{h}\neq\Omega_{h}$. This implies
$K_{h}/\Omega_{h}<1$ and

\begin{align*}
\exp\left(K_{h}/\Omega_{h}\right) & <\left(1-\frac{K_{j}}{\Omega_{h}}\right)^{-1}.
\end{align*}

This leads to

\begin{align*}
K_{h} & <\frac{1}{\Omega_{h}}\ln\Sigma_{j}^{\infty}\frac{K_{j}}{\Omega_{j}}.
\end{align*}

Since, $\frac{K_{1}}{\Omega_{1}},\frac{K_{2}}{\Omega_{2}}\ldots$ 

are positive and each are \emph{less than 1,} we get

\begin{align*}
\Sigma_{j=1}^{h}\frac{K_{j}}{\Omega_{j}} & <\left\{ \left(1-\frac{K_{1}}{\Omega_{1}}\right)\left(1-\frac{K_{2}}{\Omega_{2}}\right)\cdots\left(1-\frac{K_{h}}{\Omega_{h}}\right)\right\} ^{-1}-1
\end{align*}

(see remark \ref{remark1} in the Appendix I and also Appendix II).
Suppose, if we relax the assumption on empty classes by allowing $\eta_{h}\geqslant0$,
then $\frac{K_{h}}{\Omega_{h}}\in\left[0,1\right]$. In this case
we can use the Weierstrass inequality of the type

\begin{eqnarray}
1-\sum_{j=0}^{h}\frac{K_{h}}{\Omega_{h}} & \leq & \Pi_{j=0}^{h}\left(1-\frac{K_{h}}{\Omega_{h}}\right)\leq\left(1+\sum_{j=0}^{h}\frac{K_{h}}{\Omega_{h}}\right)^{-1}\label{2.1}
\end{eqnarray}

When $\frac{K_{h}}{\Omega_{h}}\geq0$, we have

\begin{eqnarray}
\Pi_{j=0}^{h}\left(1+\frac{K_{h}}{\Omega_{h}}\right) & \leq & \left[\left\{ h+\sum_{j=0}^{h}\frac{K_{h}}{\Omega_{h}}\right\} /h\right]<\exp\left\{ \sum_{j=0}^{h}\frac{K_{h}}{\Omega_{h}}\right\} \nonumber \\
\\
\Rightarrow\sum_{j=0}^{h}\frac{K_{h}}{\Omega_{h}} & > & \ln\left\{ \Pi_{j=0}^{h}\left(1+\frac{K_{h}}{\Omega_{h}}\right)\right\} \label{2.2}
\end{eqnarray}

(from the result by \cite{Klamkin1970}, and for $\frac{K_{h}}{\Omega_{h}}\in\left[0,1\right]$,
we have 

\begin{eqnarray}
\Pi_{j=0}^{h}\left(1-\frac{K_{h}}{\Omega_{h}}\right) & \leq & \left[\left\{ h-\sum_{j=0}^{h}\frac{K_{h}}{\Omega_{h}}\right\} /h\right]<\exp\left\{ -\sum_{j=0}^{h}\frac{K_{h}}{\Omega_{h}}\right\} \nonumber \\
\nonumber \\
\sum_{j=0}^{h}\frac{K_{h}}{\Omega_{h}} & < & \ln\left\{ \Pi_{j=0}^{h}\left(1-\frac{K_{h}}{\Omega_{h}}\right)\right\} \label{2.3}
\end{eqnarray}

Suppose $\frac{K_{j}}{\Omega_{j}}$ over $j$ for $j=1,2,...,h$ form
a probability distribution, then we can arrive at following inequality
(for details refer to \cite{Kalmkim79,El-Neweihi79}

\begin{eqnarray}
\Pi_{j=0}^{h}\left(1+\frac{K_{h}}{\Omega_{h}}\right) & \geq\frac{\left(h+1\right)^{h}}{\left(h-1\right)^{h}} & \Pi_{j=0}^{h}\left(1-\frac{K_{h}}{\Omega_{h}}\right)\label{2.4}
\end{eqnarray}

\begin{thm}
If $\Lambda_{h}>\Omega_{h}$ and MRF is present then $\Lambda_{h}\in\mathbf{B}_{\epsilon}(K_{h}-\eta_{h}).$\end{thm}
\begin{proof}
We have seen in section 2.2.1 that $\alpha<\Lambda_{h}\exp\left\{ -(K_{h}-\eta_{h}/\Lambda_{h})\right\} .$
Therefore, $\left|\Lambda_{h}-\right.$ $\left.(K_{h}-\eta_{h})\right|$
$<\Lambda_{h}\exp\left\{ -(K_{h}-\eta_{h}/\Lambda_{h})\right\} .$
When we choose

\begin{align*}
K_{h} & >\Lambda_{h}\frac{\ln\Lambda_{h}}{\ln\epsilon}+\eta_{h}
\end{align*}

for some $h>\bold{N}$ then the result follows.\end{proof}
\begin{note}
When lemma \ref{lemma1} is true then $\Lambda_{h}\in\mathbf{B}_{\epsilon}(K_{h}).$
\end{note}

\subsection{$\bold{\left(\Lambda_{h}=\Omega_{h}\right)}$}

In this situation, error of reporting is evidently null. However,
possibility of \emph{MRF} could not be ruled out. Suppose $\Omega_{h}$
is formed of $K_{h}$ classes as we saw in section 2.2.1 and $\Omega_{h}=K_{h},$
then $\alpha=0.$ If $\Omega_{h}>K_{h},$ then the arguments presented
in 2.2.1 holds here and similar error exists.

$ $

$ $

\subsection{\textmd{Stratification of error by location and time}}

Let $U$ and $V$ are $s\times t$ matrices of reported cases and
total cases across $s$ geographical locations for $t$ time points.
$U$ is represented by,

\begin{eqnarray*}
U & = & \left[\begin{array}{cccc}
\Omega_{11} & \Omega_{12} & \ldots & \Omega_{1t}\\
\Omega_{21} & \Omega_{22} & \ldots & \Omega_{2t}\\
\vdots & \vdots & \ddots & \vdots\\
\Omega_{s1} & \Omega_{s2} & \ldots & \Omega_{st}
\end{array}\right]
\end{eqnarray*}

where, $\Omega_{ij}$ is denotes the cases in $i^{th}$ location in
the $j^{th}$ time point (for $i=1,2,\cdots s$ and $j=1,2,\ldots t$).
Let $\Omega_{i.}=\Sigma_{j=1}^{t}\Omega_{ij}$ and $\Omega_{..}=\Sigma_{i=1}^{s}\Sigma_{j=1}^{t}\Omega_{ij}.$
If $\alpha_{ij}$ denote error of reporting in the $i^{th}$ location
and $j^{th}$ time point, then $V$ can be represented by,

\begin{eqnarray*}
V & = & \left[\begin{array}{cccc}
\Omega_{11}\pm\alpha_{11} & \Omega_{12}\pm\alpha_{12} & \ldots & \Omega_{1t}\pm\alpha_{1t}\\
\Omega_{21}\pm\alpha_{21} & \Omega_{22}\pm\alpha_{22} & \ldots & \Omega_{2t}\pm\alpha_{2t}\\
\vdots & \vdots & \ddots & \vdots\\
\Omega_{s1}\pm\alpha_{s1} & \Omega_{s2}\pm\alpha_{s2} & \ldots & \Omega_{st}\pm\alpha_{st}
\end{array}\right]\\
\\
\\
 & = & \left[\begin{array}{cccc}
\Lambda_{11} & \Lambda_{12} & \ldots & \Lambda_{1t}\\
\Lambda_{21} & \Lambda_{22} & \ldots & \Lambda_{2t}\\
\vdots & \vdots & \ddots & \vdots\\
\Lambda_{s1} & \Lambda_{s2} & \ldots & \Lambda_{st}
\end{array}\right]
\end{eqnarray*}

If 

\begin{eqnarray*}
\alpha_{1j}\neq0 & \textrm{and} & \alpha_{1j}=0\textrm{ \ensuremath{\forall}}j>1\\
\alpha_{2j}\neq0 & \textrm{for \ensuremath{j=1,2}and} & \alpha_{2j}=0\textrm{ \ensuremath{\forall}}j>2\\
\vdots\\
\alpha_{sj}\neq0 & \textrm{for \ensuremath{j=1,2,\ldots t}},
\end{eqnarray*}

then the characteristic roots are $\Lambda_{11},\Lambda_{22},\ldots,\Lambda_{st}.$
In the presence of an epidemic, we have, $\Omega_{11}\neq0,\Omega_{22}\neq0,\ldots,\Omega_{st}\neq0$,
hence $V$ can never be a \emph{singular}. In this situation, $V$
is always \emph{invertible}, such that:

\begin{eqnarray*}
\Lambda_{1.}= & \Sigma_{j=1}^{t}\Lambda_{1j}= & \Omega_{11}\pm\alpha_{11}\\
\Lambda_{2.}= & \Sigma_{j=1}^{t}\Lambda_{2j}= & \Sigma_{j=1}^{2}\left(\Omega_{2j}\pm\alpha_{2j}\right)\\
\vdots & \vdots & \vdots\\
\Lambda_{s.}= & \Sigma_{j=1}^{t}\Lambda_{sj}= & \Sigma_{j=1}^{t}\left(\Omega_{sj}\pm\alpha_{sj}\right)
\end{eqnarray*}

If the error in reporting cases do not follow any pattern, then the
relationship between $U$ and $V$ follow a random process. There
needs care in understanding the variability in the error, especially,
if the pandemic persists in the population for longer duration.

\section{Varying epidemic efficiency function}

We saw in the previous section that error of reporting plays important
role in understanding the epidemic even though it is taken as $\Lambda_{h}\sim\Omega_{h}$
over $h.$ Here in this section, it is assumed as a continuous random
variable with a probability density function (say $\varphi(\alpha)$).
This assumption allows variation in the error of reporting over the
time period $h$. Now the relation between total and reported cases
is taken as $\Lambda_{h}=\Omega_{h}\pm\overline{\alpha},$ where $\overline{\alpha}=\int_{-\infty}^{+\infty}\alpha\varphi(\alpha)d\alpha$
(mean reporting error).

The error of reporting might increase rapidly or stay steadily or
might decline after certain time point, since the beginning of an
epidemic. Suppose epidemic hits at time $t_{0}$, then error might
increase or decrease till $t_{k}$ and then change its direction asymptotically
(where $t_{0}<t_{k}$). The rate of increase or decrease from $t_{0}$
to $t_{k}$ could be rapidly fast or slow. To fit all such situations,
we choose Weibull and gamma functions and try to explain the error
involved through them. These tw\textbf{o }distributions can imitate
several functional forms of the nature of the error, that we are interested.
Historically, Weibull distribution has been very popular in the reliability
analysis and recently it was found to be giving satisfactory results
to model incubation period of AIDS \cite{Rao&Hira2003} and survival
distribution while analyisng bird flu data \cite{RaoASRS2008}. There
are instances where gamma distribution was also worked as a reliable
model to explain the incubation period of AIDS. These two distributions
were able to capture the variability in the incubation period\textbf{
}because of their versatile nature.\textbf{ }Suppose $\alpha\sim Weibull$
density with scale parameter $\theta$ and shape parameter $\pi$,
then the mean of the error function is $\theta\Gamma\left(1+1/\pi\right)$
and $\Lambda_{h}=\Omega_{h}\pm\theta\Gamma\left(1+1/\pi\right).$
Unless, if the reporting is extremely worst, we need not expect the
situation $\Omega_{h}<$$\:\overline{\alpha}$, hence we assume $\Omega_{h}>\theta\Gamma\left(1+1/\pi\right)\forall\, h$.
This assumption is also supported by the fact that $\alpha\sim Weibull$
implies $\alpha\rightarrow0$ ($\alpha\neq0$) as $t\rightarrow\infty.$
When $\alpha\sim gamma$ density with scale parameter $\lambda$ and
shape parameter $\nu$, then the mean of the error function is $\nu/\lambda$
and $\Lambda_{h}=\Omega_{h}\pm\nu/\lambda.$ 

When total cases exceed reported cases, \emph{MRF} discussed in the
previous section could exist. In such situation, the error estimated
above using two densities will be an under estimate\emph{.} Let $\eta$
be the factor due to \emph{MRF} which follows a Weibull density with
parameters $\left(p,q\right)$ and $\varphi(\alpha')$ be the associated
probability density function. If $\overline{\eta}$ is mean number
of empty classes out of $K_{h}$ classes, then the mean error in the
presence of \emph{MRF} is $\overline{\alpha'}$ \emph{(}say\emph{)
$=\overline{\alpha}+\overline{\eta}$.} Now, the total cases can be
estimated as $\Lambda_{h}=\Omega_{h}+p\gamma\left\{ \left(1+1/q\right),\left(K/p\right)^{q}\right\} +\theta\Gamma\left(1+1/\pi\right)$
(for \emph{Weibull}) and $\Lambda_{h}=\Omega_{h}+p\gamma\left\{ \left(1+1/q\right),\left(K/p\right)^{q}\right\} +\nu/\lambda$
(for \emph{gamma}). See \ref{remark2} in the appendix for the derivation
of $\overline{\alpha'}$. See also the difference in the mean error
among 10 pairs of $\left(\Lambda_{h},\Omega_{h}\right)$ for $\Lambda_{h}>\Omega_{h}$
situation given in the example 1. 
\begin{example}
A numerical example is given to show the difference between mean error
($\alpha$) and true mean error ($\overline{\alpha}'$) when multiple
reporting is present and $\Lambda_{h}>\Omega_{h}$.

$\begin{array}{cccccccc}
\\
\underline{\left(\Lambda_{h},\,\Omega_{h}\right)} & \underline{\alpha} & \underline{\overline{\alpha}} &  &  & \underline{\left(\Lambda'_{h},\, K_{h}-\eta_{h}\right)} & \underline{\alpha'} & \underline{\overline{\alpha}'}\\
\\
\left(100,95\right) & 5 &  &  &  & \left(100,95-40\right) & 45\\
\left(90,82\right) & 8 &  &  &  & \left(90,82-38\right) & 46\\
\left(110,100\right) & 10 &  &  &  & \left(110,\,\,100-40\right) & 50\\
\left(95,80\right) & 15 &  &  &  & \left(95,80-40\right) & 55\\
\left(102,90\right) & 12 & 7.7 &  &  & \left(102,90-35\right) & 47 & 35.7\\
\left(90,80\right) & 10 &  &  &  & \left(90,80-30\right) & 40\\
\left(117,110\right) & 7 &  &  &  & \left(117,110-20\right) & 27\\
\left(105,100\right) & 5 &  &  &  & \left(105,100-17\right) & 22\\
\left(197,194\right) & 3 &  &  &  & \left(197,194-12\right) & 15\\
\left(208,206\right) & 2 &  &  &  & \left(208,206-8\right) & 10\\
\\
\end{array}$

\emph{MRF} can be viewed as a multivariate variable and in such situation
the error estimation will be different than above. The discussion
on multivariate Weibull can be seen elsewhere \cite{Castillo90,Ahmad94}.
In these works authors have demonstrated estimation of parameters
when there are more than two parameters. 
\end{example}

\section{Conclusions}

Mathematical modeling has an important contribution in understanding
epidemic outbreak and its spread. Reporting of the infections or disease
cases are vital in terms of inputs to these models. However, at the
same time not being reported or over reporting of the cases leads
to limitations in assessing the epidemic spread. Usually, mathematical
models in epidemiology of infectious diseases consists of several
parameters, including those determine growth of an epidemic. Growth
of an epidemic at the initial stage is estimated by conducting trend
analysis of reported cases. Unless reported cases are adjusted for
under reporting (if such exists) and corresponding growth rates are
revised before plugging into models, often models need not predict
accurately the spread of infection. The difficulty lies in understanding
the degree of under reporting when a trend analysis on reported cases
is conducted. Some times reporting may be accurate in few reporting
centers but these centers might not be representative to the entire
population for which we are interested to predict the future course
of an epidemic by using mathematical models. Further, the presence
of multiple reporting within under reporting of disease cases could
complicate the assessment of degree of under reporting and hence calculation
of growth parameters required for modeling the spread is not straightforward.
In an recent outbreaks of SARS there was some concern for under reporting
\cite{Parry2003,Lange2003} and over-reporting \cite{Feng2009}, however
it was concluded later that there was no evidence of over-reporting
of SARS \cite{Liu2009}. We conclude there needs systematic adjustment
for under reporting and multiple reporting within under reporting
before analyzing the hospital based data, if such issues exists in
the data.\textbf{ }In this note, total disease cases occurred in a
given population was taken as reported plus or minus error of reporting.
We have theoretically analyzed the degree of reporting error involved
in under, over and multiple reporting of disease cases. We saw that
errors have upper bounds $\widehat{\Omega}\ln\left(\widehat{\Omega}/\widehat{\Lambda}\right)$
when $T<\Omega_{h}$ and $\widehat{\Lambda}\exp\left(-\widehat{\Omega}/\widehat{\Lambda}\right)$
when $\Lambda_{h}>\Omega_{h}.$\textbf{ }Multiple reporting factor
(MRF) influences the error estimates when disease cases are reported
more than once\textbf{ }and over all there exists under reporting
in an outbreak. We have explained schematically as well as numerically
the impact of this multiple reporting through a factor $\eta$. When
reported cases suffer from under reporting, the upper bound for error
is larger. In the presence of \emph{MRF} and $\Lambda_{h}>\Omega_{h},$
these bounds increase further. 

When the error is assumed to be a continuous random variable which
follows two probability density functions \emph{viz,} \emph{Weibull,
gamma} then the relation between total and reported cases are given
in terms of their respective means obtained from these densities.
Also, for the continuous case the impact of \emph{MRF} is studied
and error is derived using probability density functions. The error
function expressed in terms of incomplete gamma function can be numerically
explored. Such functions can also be applied for computation of bounds
of life expectancy in human populations \cite{Rao89,Harter67}. When
reported cases are completely made up of $K_{h}$ classes out of which
$\eta_{h}$ classes are empty (i.e. with no reporting in these classes)
then we showed that additional error $p\gamma\left\{ \left(1+1/q\right),\left(K/p\right)^{q}\right\} $
would be an algebraic addition to the error without \emph{MRF.} It
was also shown that as $\eta_{h}\rightarrow0,$ then $\Lambda_{h}-K_{h}\rightarrow\alpha.$
Recall, that $K_{h}$ is a positive integer defined as number of classes
at time $h$ which can accommodate\textbf{ }$\Omega_{h}$. 

In case of emerging or newly identified pandemics, reporting error
could follow a random pattern. Sometimes, the reporting across countries
also vary in case of new epidemics due to lack of proper guidelines
and protocols of diagnosis. The matrix analysis presented can be extended
to global epidemic, where status of error in each country is depended
on the country specific guidelines. The results presented in this
work helps in framing protocols for analysis and reporting the epidemic
data. The results can be useful in careful handling of various factors
of potential errors due to multiple reporting independently and multiple
reporting within under reporting. This kind of analysis presented
here applied to the epidemic is new and probably is in initial stage.
We are able to address the issues related to importance of adjusting
multiple reporting error by this method. The ideas presented could
lead to new theoretical approaches and also could be a supplement
to the existing methods in epidemic analysis.

\section*{Appendix I}
\begin{rem}
\label{remark1}Suppose $0<\frac{K_{j}}{\Omega_{j}}<1$ $\forall\, j=1,2,\cdots,h$.
Then

\begin{eqnarray*}
\left(1+\frac{K_{1}}{\Omega_{1}}\right)\left(1+\frac{K_{2}}{\Omega_{2}}\right) & > & 1+\left(\frac{K_{1}}{\Omega_{1}}+\frac{K_{2}}{\Omega_{2}}\right)\\
\left(1+\frac{K_{1}}{\Omega_{1}}\right)\left(1+\frac{K_{2}}{\Omega_{2}}\right)\left(1+\frac{K_{3}}{\Omega_{3}}\right) & > & 1+\left(\frac{K_{1}}{\Omega_{1}}+\frac{K_{2}}{\Omega_{2}}+\frac{K_{3}}{\Omega_{3}}\right)\\
 & \vdots\\
 &  & \textrm{and so on up to \ensuremath{h\, th}term.}
\end{eqnarray*}

Therefore, we get 
\begin{eqnarray*}
\Sigma_{j=1}^{h}\frac{K_{j}}{\Omega_{j}} & < & \prod_{j=1}^{h}\left(1+\frac{K_{j}}{\Omega_{j}}\right)-1\\
 & < & \left\{ \prod_{j=1}^{h}\left(1-\frac{K_{j}}{\Omega_{j}}\right)\right\} ^{-1}-1
\end{eqnarray*}

\end{rem}
This kind of inequality is also called Weierstrass's type inequality.
Original inequality is given in Appendix II. 
\begin{rem}
\label{remark2}Let $\alpha$ and $\eta$ be two continuous random
variables with $0<\alpha<\infty$ and $0<\eta<K$, where $K$ is the
maximum number of empty classes that $\eta$ can attain. We know that
$E(\overline{\alpha'})=E(\overline{\alpha})+E(\eta),\textrm{ where \ensuremath{E}is expectation }$
$\textrm{or mean of the random variable.}$ This means, $\overline{\alpha'}=\overline{\alpha}+\overline{\eta}.$
Let $\,\alpha\sim Weibull\,(\theta,\pi)$ and $\eta\sim Weibull\,(p,q)$
then 

\begin{eqnarray}
\overline{\alpha'} & = & \frac{\pi}{\theta}\int_{0}^{\infty}\alpha\left(\frac{\alpha}{\theta}\right)^{\pi-1}\exp\left\{ -\left(\frac{\alpha}{\theta}\right)^{\pi}\right\} d\alpha+\frac{q}{p}\int_{0}^{K}\eta\left(\frac{\eta}{p}\right)^{q-1}\exp\left\{ -\left(\frac{\eta}{p}\right)^{q}\right\} d\eta\label{A1}
\end{eqnarray}

Taking $\left(\frac{\alpha}{\theta}\right)^{\pi}=w,$ and $\left(\frac{\eta}{p}\right)^{q}=u$
and changing the limits accordingly, we get as below

\begin{eqnarray*}
 & = & \theta\Gamma\left(1+\frac{1}{\pi}\right)+\frac{q}{p}\int_{0}^{\left(K/p\right)^{q}}pu^{\frac{1}{q}}\frac{up}{pu^{1/q}}\exp\left(-u\right)\left(\frac{p}{q}\right)u^{\frac{1}{q}-1}du\\
 & = & \theta\Gamma\left(1+\frac{1}{\pi}\right)+p\int_{0}^{\left(K/p\right)^{q}}u^{\frac{1}{q}}\exp\left(-u\right)du\\
\overline{\alpha'} & = & \theta\Gamma\left(1+\frac{1}{\pi}\right)+p\gamma\left\{ \left(1+1/q\right),\left(\frac{K}{p}\right)^{q}\right\} 
\end{eqnarray*}
\end{rem}
\begin{note}
Other possible assumptions like $\eta\sim gamma(p,q)$ and derivation
of corresponding mean error is left as an exercise.
\end{note}

\section*{Appendix II: Results due to Copson \cite{Copson70/71}, Klamkin and
Newman \cite{Klamkin1970}, Klamkin \cite{Kalmkim75}, El-Neweihi
and Proschan \cite{El-Neweihi79} }

E.T. Copson proved that a bounded sequence of real numbers $\left(a_{n}\right)$
is convergent if the inequality $a_{n+2}\leq\frac{1}{2}\left(a_{n+1}+a_{n}\right)$
is satisfied.

He further proves a more general theorem, whose statement is as follows:
\begin{thm*}
If $\left(a_{n}\right)$ is a bounded sequence which satisfies the
inequality $a_{n+r}=\sum_{r=1}^{s}K_{s}a_{n+r-s}$, where the coefficients
$K_{s}$ are strictly positive and $K_{1}$$+K_{2}$ $+...+K_{r}$$=1$,
then $\left(a_{n}\right)$ is a convergent sequence. But if $\left(a_{n}\right)$
is unbounded, it diverges to $-\infty.$
\end{thm*}
Weierstrass inequalities \cite{Bromwich1965} (also available in \cite{Klamkin1970})
are given by

\begin{align}
1-S_{1} & \leq\prod_{i=1}^{n}\left(1-A_{i}\right)\leq\left(1+S_{1}\right)^{-1}\label{eq:weir1}\\
1+S_{1} & \leq\prod_{i=1}^{n}\left(1+A_{i}\right)\leq\left(1-S_{1}\right)^{-1}\label{eq:weir2}
\end{align}

where $A_{1},$ $A_{2},$ $...,A_{n}$ are real numbers in $[0,1]$
and $S_{1}=\sum_{i=1}^{n}A_{1}.$ $S_{1}\leq1$ in the inequality
(\ref{eq:weir2}). M. S. Klamkin and D. J. Newman \cite{Klamkin1970}
have extended the Weierstrass inequalities and showed that

\begin{align}
\prod_{i=1}^{n}\left(1+A_{i}\right)\geq & \left(n+1\right)^{n}\prod_{i=1}^{n}A_{i}\label{eq:klam_new1}\\
\prod_{i=1}^{n}\left(1-A_{i}\right)\geq & \left(n-1\right)^{n}\prod_{i=1}^{n}A_{i}\label{eq:klam-new2}
\end{align}

where $A_{i}\geq0$, $i=1,2,...n$ and $\sum_{i=1}^{n}A_{i}=1.$ M.
S. Klamkin \cite{Kalmkim75} further proved, under the same conditions,
that

\begin{align}
\frac{\prod_{i=1}^{n}\left(1+A_{i}\right)}{\left(n+1\right)^{n}}\geq & \frac{\prod_{i=1}^{n}\left(1-A_{i}\right)}{\left(n-1\right)^{n}}\label{eq:klam1}
\end{align}

with equality only if $A_{i}=1/n$. E. El-Neweihi and F. Proschan
\cite{El-Neweihi79} had established Weierstrass-product type inequalities
(\ref{eq:klam_new1}, \ref{eq:klam-new2}, \ref{eq:klam1}) by a uniform
approach and then using powerful tools of majorization and Schur-convex
and Schur-concave functions. 

\pagebreak

\begin{table}
\begin{tabular}{|c|c|}
\hline 
Parameter & Definition\tabularnewline
\hline 
\hline 
$\Lambda_{h}$ & number of total disease cases at time $h$\tabularnewline
\hline 
$\Omega_{h}$ & number of reported disease cases at time $h$ \tabularnewline
\hline 
$\lambda_{m}$ & $\sum_{k=1}^{m}\Lambda_{k}$, where $\Lambda_{k}$ is number of total
disease cases at time $h$ \tabularnewline
\hline 
$\omega_{m}$ & $\sum_{k=1}^{m}\Omega_{k}$, where $\Omega_{k}$is number of reported
disease cases at time $h$ \tabularnewline
\hline 
$n_{1h}$ & number of individuals out of $\Omega_{h}$ who are reported exactly
once\tabularnewline
\hline 
$\alpha_{h}$ & difference between $\Omega_{h}$ and $\Lambda_{h}$\tabularnewline
\hline 
$\alpha'_{h}$ & difference between $\Lambda_{h}$and $n_{1h}$ , where $n_{1h}<\Omega_{h}$\tabularnewline
\hline 
$K_{h}$ & number of classes where $\Omega_{h}$cases could be located\tabularnewline
\hline 
$\eta_{h}$ & number of empty classes out of $K_{h}$\tabularnewline
\hline 
\end{tabular}

\caption{Parameters and definitions}

\end{table}

$ $

\pagebreak
\end{document}